\documentstyle[preprint,aps]{revtex}
\begin{document}
\draft
\title{Non-linear waves in Fermi liquids}
\author{A.~H.~Castro Neto}
\bigskip
\address
{Institute for Theoretical Physics\\
University of California at Santa Barbara\\
Santa Barbara, CA, 93106-4030}

\maketitle

\begin{abstract}
I show that when non-linearities are taken into account
the Landau theory of Fermi liquids predicts
the existence of hyperbolic waves in
fermionic systems. The zero sound is described
by a infinite set of coupled non-linear partial differential equations,
one for each harmonic of oscillation of the Fermi surface.
When just a few harmonics are included it is possible to describe
the dynamics of the first sound in a very simple way.
These results lead to the interesting experimental possibility
of having non-linear effects in quantum liquids. In particular
I discuss the problem of generation of second harmonic in a
Fermi system.
\end{abstract}

\bigskip


\narrowtext

It is already well established that the dynamics of sound waves in
neutral systems such as $He^3$ or plasmons in charged systems can be
well described
by the Landau theory of Fermi liquids \cite{baym}. The usual approach
is based on the linearized Boltzmann equation from which
response functions and the dispersion of the collective modes
are obtained.
It is known that the same type of linearized equation
leads to the correct description of the low energy excitations
of the Tomonaga-Luttinger model in one dimension \cite{tomonaga}.
This result means that in one dimension, due to the lack of a
particle-hole continuum at low energies and long wavelengths,
the bosonic excitations are simply sound waves. It has been shown
recently by the generalization of the bosonization procedure in
higher dimensions that the actual bosonic excitations of fermionic systems
at low energies are the displacements of the Fermi surface, that
is, particle-hole continuum plus collective modes \cite{bosonization}.
Thus although the quasiparticle residue is zero in an one
dimensional model with interactions, the Landau expansion for the
energy and the Boltzmann equation are still valid.

Amazingly enough, the Landau theory
is capable of predicting the existence of a first order phase
transition. This transition happens at the point where the
compressibility vanishes (spinodal line) and the system
becomes unstable to long wavelength fluctuations. The
Landau theory predicts that there is an exponential
growth of the collective mode inside the many-body system leading
to phase separation. This is also known as Pomeranchuk's instability
\cite{pomeranchuk} and has been used to study phase separation
in mixtures of helium, processes involving heavy ion collisions
at high energies \cite{pethick} and more recently,
the problem of phase separation in two dimensional systems such as
some cuprates and nicklates \cite{kivelson,separation}.

The main result of the linearization of the Boltzmann equation
is that it leads to plane waves which propagate in space with
a dispersion defined by the poles of the density-density response
function \cite{baym}. In the case of instabilities the frequency
of oscillation of the wave is found to be imaginary leading to
an exponential growth of the amplitude \cite{pethick}.
In principle, for short range interactions, the growth in unbounded.
However, in real systems
we always have long range forces (such as the Coulomb force) or
dissipative mechanisms which
tend to inhibit the growth of the wave leading to saturation of
the wave amplitude \cite{separation}.

I am going to show that, within the framework of Landau theory,
if non-linear terms are taking into account, the physics of the
waves which propagate in Fermi systems
is not described by ordinary plane waves but by the so-called
hyperbolic waves \cite{shock}. The non-linearities
appear when we consider fluctuations of the Fermi surface which
are comparable to the Fermi momentum. This is similar to
the problem of the inclusion of large momentum scattering in
one dimension (backscattering or umklapp) which leads to
Sine-Gordon theories \cite{emery}. The terms that we are going
to keep shall lead to non-linear bosonic theories in dimensions
higher than one \cite{bosonization}.

I am going to
review the main points on the physics of waves in Landau theory.
In this way it will become clear the origin of the non-linearities
I am talking about. The starting point is the Landau theory which
assumes that the local change in energy due to
a local change in the quasiparticle distribution,
$\delta n(\vec{k},\vec{r},t)$, is given by ,
\begin{equation}
\Delta E (\vec{r},t)=
\sum_{\vec{k}} \left(\epsilon^0_{k}+U(\vec{r},t)\right)
\, \, \delta n(\vec{k},\vec{r},t) +
\frac{1}{V} \sum_{\vec{k},\vec{k}'} f_{\vec{k},\vec{k}'} \,
\delta n(\vec{k}',\vec{r},t) \, \, \delta n(\vec{k},\vec{r},t)
\end{equation}
where $ \epsilon^0_{k}$ is the bare dispersion relation,
$U(\vec{r},t)$ is an external potential applied to the system,
$V$ is the volume and $f_{\vec{k},\vec{k}'}$ is the quasiparticle
two-body interaction (for simplicity I am going to assume that
these interactions are homogeneous, that is, do not depend on the
position of the quasiparticles).

Then the actual  quasiparticle dispersion is obtained from (1),
\begin{equation}
\epsilon_{\vec{k}}(\vec{r},t) = \epsilon^0_{k} + U(\vec{r},t)+
\frac{1}{V} \sum_{\vec{k}'} f_{\vec{k},\vec{k}'} \, \,
\delta n(\vec{k}',\vec{r},t).
\end{equation}
It will become clear that non-linearities appear
because the quasiparticle dispersion depends on the quasiparticle
distribution. This is equivalent to problem of flow of fluids
where the velocity of propagation depends on the density.
At this level the equations of motion for the quasiparticles
are derived from (2) as,
\begin{eqnarray}
\frac{\partial \vec{r}}{\partial t} &=& \nabla_{\vec{k}}
\epsilon_{\vec{k}}(\vec{r},t)
\nonumber
\\
\frac{\partial \vec{k}}{\partial t} &=& - \nabla_{\vec{r}}
\epsilon_{\vec{k}}(\vec{r},t).
\end{eqnarray}

The main problem now is how to describe the evolution in time of
the quasiparticle distribution given the equations (3).
In the Landau theory the quasiparticle distribution, $n(\vec{k},\vec{r},t)$,
changes because the scattering of quasiparticles and it is
described by the Boltzmann equation,
\begin{eqnarray}
\frac{d n(\vec{k},\vec{r},t)}{d t} = I(n)
\end{eqnarray}
where $I(n)$ is the collision integral. Eq.(4) is written in
terms of the partial derivatives with help of (3) as,
\begin{equation}
\frac{\partial n(\vec{k},\vec{r},t)}{\partial t}
+ \nabla_{\vec{r}}n(\vec{k},\vec{r},t) \cdot \nabla_{\vec{k}}
\epsilon_{\vec{k}}(\vec{r},t) -
\nabla_{\vec{k}}n(\vec{k},\vec{r},t) \cdot \nabla_{\vec{r}}
\epsilon_{\vec{k}}(\vec{r},t) = I(n).
\end{equation}

As usual we expand the distribution around its equilibrium value,
that is,
\begin{eqnarray}
n(\vec{k},\vec{r},t) = n_0(k) + \delta n(\vec{k},\vec{r},t)
\end{eqnarray}
where $n_0(k) = \Theta(\mu-\epsilon^0_{k})$ with $\mu$ the chemical
potential. Substituting (2) and (6) in (5) and using that
$\nabla_{\vec{k}} \epsilon^0_{k}= \vec{v}_{k}$ and
$\nabla_{\vec{k}} n_0(k) =- \vec{v}_{k} \delta(\mu-\epsilon^0_{k})$
we easily find,
\begin{eqnarray}
\frac{\partial \delta n(\vec{k},\vec{r},t)}{\partial t} &+&
 \vec{v}_{k} \cdot \nabla_{\vec{r}} \delta n(\vec{k},\vec{r},t)
+\delta(\mu-\epsilon^0_{k}) \left(\vec{v}_{\vec{k}} \cdot \nabla_{\vec{r}}
U(\vec{r},t) +\frac{1}{V} \sum_{\vec{k}'}
f_{\vec{k},\vec{k}'} \, \, \vec{v}_{k} \cdot \nabla_{\vec{r}}
\delta n(\vec{k}',\vec{r},t)\right)
\nonumber
\\
&+&\nabla_{\vec{k}} \delta n(\vec{k},\vec{r},t) \cdot \nabla_{\vec{r}}
U(\vec{r},t)+
\frac{1}{V} \sum_{\vec{k}'} \left(\nabla_{\vec{k}} f_{\vec{k},\vec{k}'}
\cdot \nabla_{\vec{r}} \delta \, \, n(\vec{k},\vec{r},t) \right)
\delta n(\vec{k}',\vec{r},t)
\nonumber
\\
&-&\frac{1}{V} \sum_{\vec{k}'} f_{\vec{k},\vec{k}'} \, \,
\nabla_{\vec{k}} \delta n(\vec{k},\vec{r},t) \cdot
  \nabla_{\vec{r}} \delta n(\vec{k}',\vec{r},t) = I(n) .
\end{eqnarray}
I have kept all terms in the equation and I am not assuming that
the deviations from equilibrium are small.
Observe that the last two terms on the l.h.s. of (7)
are not present in the usual
approach of the Fermi liquid theory \cite{baym,pines}
because they are second order in the deviations. These terms give rise to
non-linear effects as I will show.

Since it is expected that the main physics of the problem
depends only on the dynamics {\it at} the Fermi surface
(which is usually the case in fermionic systems in any
number of dimensions) the solution of (7) can be written as,
\begin{equation}
\delta n(\vec{k},\vec{r},t) = \delta(\mu-\epsilon^0_{\vec{k}})
\rho(\vec{k},\vec{r},t)
\end{equation}
where $\rho(\vec{k},\vec{r},t)$ obeys the following equation,
\begin{eqnarray}
\frac{\partial \rho(\vec{k},\vec{r},t)}{\partial t} &+&
 \vec{v}_{k} \cdot \left(\nabla_{\vec{r}} U(\vec{r},t)+
\nabla_{\vec{r}} \rho(\vec{k},\vec{r},t)\right)
+\frac{1}{V}  \sum^{'}_{\vec{k}'} f_{\vec{k},\vec{k}'} \vec{v}_{k} \cdot
\nabla_{\vec{r}} \rho(\vec{k}',\vec{r},t)
\nonumber
\\
&+& \frac{1}{V} \sum^{'}_{\vec{k}'} \left(
\nabla_{\vec{k}} f_{\vec{k},\vec{k}'}
\cdot \nabla_{\vec{r}} \rho(\vec{k},\vec{r},t)\right)
\rho(\vec{k}',\vec{r},t)
\nonumber
\\
&+&\frac{1}{V} \sum^{'}_{\vec{k}'} \left(\nabla_{\vec{k}} f_{\vec{k},\vec{k}'}
\cdot \nabla_{\vec{r}} \rho(\vec{k}',\vec{r},t)\right)
\rho(\vec{k},\vec{r},t) = \tilde{I}( \rho ),
\end{eqnarray}
where I have used $\rho_{\vec{k}} f_{\vec{k},\vec{k}'} \nabla_{\vec{k}}
\delta(\mu-\epsilon^0_{k}) = - \delta(\mu-\epsilon^0_{k})
\nabla_{\vec{k}} (\rho_{\vec{k}} f_{\vec{k},\vec{k}'})$ and
defined $\sum^{'}_{\vec{k}'} = \sum_{\vec{k}'} \delta(\mu-\epsilon^0_{k'})$.

For simplicity I assume a spherical Fermi surface. In this way
I can simplify further the
problem by expanding $\rho_{\vec{k}}$ in Legendre polynomials in
three dimensions (a completely similar procedure can be done in
two dimensions in terms of Fourier series \cite{separation}),
\begin{equation}
\rho(\theta,\vec{r},t) = v_F \sum_{l=0}^{\infty}
u_l(\vec{r},t) \, \, P_l(\cos \theta)
\end{equation}
where $\theta$ is the angle on
the Fermi surface that parameterizes $\vec{k}$ and $v_F$ is
the Fermi velocity. Analogously the
quasiparticle interaction can also be expanded as,
\begin{equation}
f_{\vec{k},\vec{k}'}=\sum_{l=0}^{\infty}
f_l \, \, P_l(\cos \theta_{\vec{k},\vec{k'}}).
\end{equation}
Since in this paper I will consider the problem of instabilities and
non-linearities in Fermi liquid theory I will keep only the two
first harmonics in the interaction. Firstly it can be seen from (9) that
in order to have non-linearities in the problem it is necessary to consider
an interaction which has dispersion at the Fermi surface ($\nabla_{\vec{k}}
f_{\vec{k},\vec{k}'} \neq 0$) that is, at least $f_1$ must be different
from zero.
Also, due to Galilean invariance \cite{baym,pines}, the effective mass of the
quasiparticles depends only on $f_1$, namely,
 $\frac{m^*}{m}=1 + \frac{N(0) f_1}{3}$,
where $N(0) = \frac{1}{V} \sum_{\vec{k}}
\delta(\mu-\epsilon^0_{k}) =\frac{m^* k_F}{\pi^2} $ is the density of
states at the Fermi surface. Secondly, the compressibility of a
Fermi liquid,
$\kappa$, depend on the parameter $f_0$ \cite{baym} ($\kappa =
\frac{N(0)}{n^2(1+N(0)f_0)}$ where $n$ is the density) and
the instabilities to phase separation, or Pomeranchuk's
instabilities, appear for $N(0) f_0 < -1$ when the systems
crosses the spinodal line \cite{pethick}. Thus, in this
approach it will be sufficient to consider only the first
two harmonics, that is,
\begin{eqnarray}
F(\vec{k},\vec{k}') = F_0 + F_1 \cos(\theta_{\vec{k},\vec{k'}})
\end{eqnarray}
where $F_{\vec{k},\vec{k}'} = N(0) f_{\vec{k},\vec{k}'}$.

Notice that in eq. (9) we just need the following summation,
\begin{equation}
\frac{1}{V} \sum^{'}_{\vec{k}'} f_{\vec{k},\vec{k}'} \rho_{\vec{k}'}
= v_F \left(F_0 u_0 + \frac{1}{3} F_1 u_1 \cos \theta \right),
\end{equation}
which will be used later in the paper.

In momentum space the Fermi velocity and momentum are radial and written as,
$\vec{v}_{k} = v_F \hat{r}$ and $\vec{k} = k_F \hat{r}$,
where $v_F$ and $k_F$ are the Fermi velocity and momentum,
respectively. For simplicity I will assume that the external
potential depends only on the $z$ direction in real space, that is,
$U(\vec{r},t) = U(z,t)$. Thus it is natural to choose a reference
frame in real space
where the axis coincide with the axis in the momentum space, that is,
$\hat{r} = \cos \theta \hat{z} + \sin \theta \,  \cos \phi \hat{x}
+\sin \theta \, \sin \phi \hat{y}$, $\hat{\theta} = -\sin \theta \hat{z}
+ \cos \theta \, \cos \phi \hat{x}+ \cos \theta \, \sin \phi \hat{y}$
and $\hat{\phi} = -\sin \phi \hat{x}+\cos \phi \hat{y}$ where
$\phi$ is the azimuthal angle.
Thus the gradient in momentum space is simply
$\nabla_{\vec{k}} = \frac{1}{k_F} \hat{\theta} \frac{\partial}{\partial
\theta}+\frac{\hat{\phi}}{k_F \sin \theta} \frac{\partial}{\partial \phi}$.
With this choice the propagation is one dimensional.
Of course, more general frames can be chosen but this simple choice
will help us to understand the dynamics of this system.
Using the above expressions, from (9) and (13), one finds,
\begin{eqnarray}
\frac{\partial \rho(\theta,z,t)}{\partial t} +
v_F \cos \theta \left(\frac{\partial U(z,t)}{\partial z} +
\frac{\partial \rho(\theta,z,t)}{\partial z}\right)+
\nonumber
\\
+ v^2_F \left[ F_0 \, \cos \theta \frac{\partial u_0(z,t) }{\partial z}
+ \frac{1}{3} F_1 \cos^2 \theta\frac{\partial u_1(z,t)}{\partial z}
\right]
\nonumber
\\
+\frac{F_1 v_F}{3 k_F} \sin^2 \theta \frac{\partial (u_1(z,t)
\rho(\theta,z,t))}{\partial z}=I(u)
\end{eqnarray}
which is the final set of equations which must be solved in order
to get the dynamics of the system.
Observe that the non-linearity (last
term in the l.h.s. of (14)) comes with
higher powers in $k^{-1}_F$ which shows that this term is important for
fluctuations out of equilibrium. It is also easy to see that by projecting the
above equation in each of its Legendre components the final set
of coupled partial differential equations has the form ($U=0$),
\begin{equation}
\frac{\partial u_n(z,t)}{\partial t} + \sum_{m=0}^{\infty}
C_{n m}(\{u\}) \frac{\partial u_m(z,t)}{\partial z} = \tilde{I}(u)
\end{equation}
where $C_{n m}(\{u\})$ is a velocity matrix which depends on the
components $u_n$. This set describes a system of coupled hyperbolic
waves (and, in particular, shock waves) where the damping depends now on the
collision integral $I(u)$. Notice that this is a general result
of the Landau theory which is obtained just from the fact that
the quasiparticle dispersion depends on the quasiparticle
distribution, eq.(2).

In order to understand and simplify the problem
I consider the hydrodynamic limit where the collisions dominate,
that is, when it is possible to have local thermal equilibrium in
the system. In this case the sound mode that propagates is the
first sound \cite{baym}. The dispersion of the sound mode can
be obtained in Landau theory straight from the compressibility,
$\kappa$ \cite{baym,separation}, and it has the dispersion,
\begin{eqnarray}
\omega_{q} &=& c_s q
\nonumber
\\
c_s &=& \sqrt{\frac{(1+F_0) \left(1+\frac{F_1}{3}\right)}{3}} v_F.
\end{eqnarray}
The first sound propagates like a plane wave and at the spinodal
line, that is, when $F_0 < -1$, the amplitude of the wave
grows exponentially given rise phase separation.
In the hydrodynamic limit, since the collisions dominate,
the quasiparticles and collective modes decay. In this case
it is sufficient to consider only the two first Fourier
components in (10), that is,
\begin{equation}
\rho(\theta,z,t) \approx v_F \left[u_0(z,t) + u_1(z,t) \cos \theta
\right] .
\end{equation}
It is worth mention that in the collisionless limit, when the
sound mode is the zero sound, it is necessary to solve the
complete set (14), or at least, to calculate the {\it non-linear}
response function of the system \cite{preparation}.

By substitution of (17) in (14) and projecting into the Legendre
components we find the following set of coupled
equations,
\begin{eqnarray}
\frac{\partial u_0}{\partial t} &+& \frac{2 F_1 v_F}{9 k_F} u_1
\frac{\partial u_0}{\partial z} +
\left[\frac{v_F}{3} \left(1+\frac{F_1}{3}\right)+
\frac{2 F_1 v_F}{9 k_F} u_0\right]
\frac{\partial u_1}{\partial z} = 0
\nonumber
\\
\frac{\partial u_1}{\partial t} &+& \frac{4 F_1 v_F}{9 k_F} u_1 \frac{\partial
u_1}{\partial z} + v_F \, (1+F_0)
\frac{\partial u_0}{\partial z} = - \frac{\partial U}{\partial z}
- \frac{u_1}{\tau},
\end{eqnarray}
where I have introduced in the place of the collision integral the
relaxation time $\tau$ in the second equation. This term is
absent in the first equation because of the conservation of
number of particles. Here I am going to assume that
this term is independent of the position. In some
cases we can consider $\tau$ to be momentum dependent which
will lead to some interesting physics.
Notice that (18) resembles the Navier-Stokes equation
for a fluid with a frictional term \cite{shock}. If $\tau^{-1}(\vec{r})=
\nu \nabla_{\vec{r}}^2$ where $\nu$ is the viscosity of the
quantum fluid then we end up with a viscous
term exactly as in the Navier-Stokes case. Already in this form, eq.(18) has
an incredibly large number of interesting physical applications
and can be studied in different ways. It would be interesting to
apply the methods of critical phenomena to understand phase transitions
in fermionic systems via renormalization group methods \cite{hohenberg,nigel}
and relate these equations to the problem of formation of patterns
\cite{cross}. It is also worth to point out that if we use a viscous
term in the equations we can define a {\it Reynolds number}, $R =
c_s L/\nu$ where $L$ is some characteristic length of the system
and we can talk about {\it turbulence} in Fermi liquids.

Observe that for $k_F \to \infty$ (18) reduces to to the
usual linear problem,
\begin{eqnarray}
\frac{\partial \tilde{u}_0}{\partial t} &+&
\frac{v_F}{3} \left(1+\frac{F_1}{3}\right)
\frac{\partial \tilde{u}_1}{\partial z} = 0
\nonumber
\\
\frac{\partial \tilde{u}_1}{\partial t} &+& v_F (1+F_0)
\frac{\partial \tilde{u}_0}{\partial z}
= -\frac{\partial U}{\partial z}
-\frac{u_1}{\tau}
\end{eqnarray}
which can be solved by Fourier transform,
\begin{eqnarray}
\tilde{u}_0(q,\omega) &=&
\frac{v_F}{3} \left(1+\frac{F_1}{3}\right) \frac{q^2 \, {\cal U}(q,\omega)}{
\omega^2-c_s^2 q^2 -i \frac{\omega}{\tau}}
\nonumber
\\
\tilde{u}_1(q,\omega) &=&
\frac{q \, \omega \, {\cal U}(q,\omega)}{
\omega^2-c_s^2 q^2 -i \frac{\omega}{\tau}}
\end{eqnarray}
where $c_s$ is the sound velocity, given in (16),
and $ {\cal U}(q,\omega)$ is the Fourier transform of $U(z,t)$.
Observe that the poles of the solutions above, that is, the eigenfrequencies
of the problem are given by,
\begin{equation}
\omega^{\pm}_q = -\frac{i}{2 \tau} \pm
\sqrt{\frac{(1+F_0)(1+\frac{F_1}{3})}{3} v_F^2 q^2-\frac{1}{4 \tau^2}}.
\end{equation}
When $\tau \to \infty$ the spectrum is the one expected
for a hydrodynamic sound, eq.(16), there is no dissipation and the
sound propagates as a plane wave. If $\tau$ is finite and $F_0 > -1$
($\kappa >0$) the the roots are pure imaginary numbers
for $q \geq (\tau v_F \sqrt{4/3 (1+F_0) (1+F_1/3)})^{-1}$, that is,
$\omega^{\pm}_q = -i \Gamma^{\pm}(q)$ where $\Gamma^{\pm}(q)$ is
a positive real number. It means that the plane wave disappears exponentially,
that is, impurities kill the sound mode which is overdamped.
For $F_0 > -1$ and $q \le (\tau v_F \sqrt{4/3 (1+F_0) (1+F_1/3)})^{-1}$
the sound mode is underdamped. However, if $F_0 < -1$
($\kappa <0$) the roots are purely imaginary and
amplitude of the wave grows exponentially in time (in a time scale
is $(\Gamma^{-}(q))^{-1}$). That is, impurities alone cannot stop
the phase separation unless $\tau$ depends on the momentum $q$.
However with a long range Coulomb repulsion one
changes $F_0 \to F_0 + \frac{2 \pi N(0) e^2}{\epsilon q}$ and
the solution of the wave equation
grows exponentially only for momentum transfer larger than
$\sqrt{\frac{4 e^2 m k_F (1+F_1/3)}
{\pi \epsilon |1+F_0|}}$. That is, there is a maximum size for the formation
of an unstable droplet \cite{separation}.
Once the instability develops in the system, the set (19)
is not valid anymore and it is necessary to use the complete set (18) and
consider non-linear effects.

In order to understand these effects I will consider the response
of this system to a small perturbation $U$. I am going to write
an expansion for $u_0$ and $u_1$ as,
\begin{eqnarray}
u_0(z,t) = \tilde{u}_0(z,t) + \delta u_0(z,t)
\nonumber
\\
u_1(z,t) = \tilde{u}_1(z,t) + \delta u_1(z,t)
\end{eqnarray}
where $\tilde{u}_0(z,t)$ and $\tilde{u}_1(z,t)$ are given in (20)
and are the {\it linear response} to the applied external potential.
$\delta u_0(z,t)$ and $\delta u_1(z,t)$ are the {\it non-linear
response} of the system to $U$. In particular we are interested in
the case where they are proportional to $U^2$. By substitution of
(22) in (18) and using (19) we easily find that,
\begin{eqnarray}
\frac{\partial \delta u_0(z,t)}{\partial t} &+& \frac{v_F}{3}
\left(1+\frac{F_1}{3}\right) \frac{\partial \delta u_1(z,t)}{\partial z}
= - \frac{2 F_1 v_F}{9 k_F} \frac{\partial
\tilde{u}_0(z,t) \tilde{u}_1(z,t)}{\partial z}
\nonumber
\\
\frac{\partial \delta u_1(z,t)}{\partial t} &+& v_F (1+F_0)
\frac{\partial \delta u_0(z,t)}{\partial z} =
- \frac{2 F_1 v_F}{9 k_F} \frac{\partial
\tilde{u}^2_1(z,t)}{\partial z}
\end{eqnarray}
these equations can be solved by Fourier transform.

Suppose, for simplicity, that the external potential is monochromatic
with some characteristic frequency $\Omega$ and wave vector $p$,
that is, ${\cal U}(k,\omega) = U_0 \delta(k-p) \delta(\omega-
\Omega)$). Then we have,
\begin{eqnarray}
\tilde{u}_0(k,\omega) &=& \frac{c_s^2 U_0}{v_F(1+F_0)} \, \,
\frac{p^2}{\Omega^2-c_s^2 p^2-i\frac{\Omega}{\tau}} \, \,
\delta(k-p) \delta(\omega-\Omega)
\nonumber
\\
\delta u_0(k,\omega) &=& \frac{F_1 c_s^2 U_0^2}{9 \pi^2 (1+F_0) k_F}
\, \, \frac{p^2 \Omega^2}{\left(\Omega^2 - c_s^2 p^2\right)
\left(\Omega^2-c_s^2 p^-i\frac{\Omega}{\tau}\right)^2}
\, \, \delta(k-2 p) \delta(\omega-2 \Omega)
\end{eqnarray}
which means that the second harmonic is generated in the system.
Thus a second peak must appear in an experiment which
probes density fluctuations.
The strength of this peak is proportional to the square of the
strength of the imposed external potential. Observe that the
ratio between the
height of the peak of the first harmonic to the height of
the second harmonic is given by,
\begin{eqnarray}
\frac{\tilde{u}_0(p,\Omega)}{\delta u_0(2 p,2 \Omega)}=
\frac{m^*-m}{3 \pi^2 m m^*} \frac{p^2 \Omega^2}{\left(\Omega^2-
c_s^2 p^2\right) \left(\Omega^2-c_s^2 p^2 -i \frac{\Omega}{\tau}
\right)} \, \, U_0
\end{eqnarray}
Experiments in $He^3$ have dealt only with linear effects which
are described in (20) \cite{wheatley}. As far as I know, there
have been no measurements of these interesting non-linear
effects in bulk $He^3$. In particular it would be interesting
to verify the existence of the second harmonic generation in
$He^3$. The same effect was observed more than thirty years
ago in non-linear optical systems \cite{franken}.

I am deeply indebted to N. Goldenfeld for asking me the right questions
and J.Langer, L. Radzihovsky and A.Zee for many
illuminating comments and suggestions. I also
acknowledge L.-Y. Chen, M.C. Cross,
L. Balents, G.Baym, D.K.Campbell, M.P.A. Fisher
and A.Tikofsky for many discussions.
This research was supported in part by the National Science Foundation
under the Grant No. PHY89-04035.

\end{document}